\documentclass[12pt]{article}
\usepackage{mathrsfs}
\usepackage{amssymb}
\usepackage{textcomp}
\usepackage[numbers]{natbib}
\usepackage{float}
\usepackage{CJK}
\usepackage[dvips]{graphics}
\usepackage[sumlimits]{amsmath}
\usepackage{pifont}
\usepackage{bbding}
\usepackage{color}
\usepackage{calrsfs}
\usepackage{bbm}
\usepackage{amssymb}
\usepackage{amsmath}
\usepackage{amsfonts}
\usepackage{multirow}
\usepackage{tikz}
\usetikzlibrary{arrows}
\usetikzlibrary{decorations.pathmorphing} 
\usetikzlibrary{positioning}
\usetikzlibrary{fit}					
\usetikzlibrary{backgrounds}	
\usetikzlibrary{shapes,snakes,calendar,matrix,backgrounds,folding}
\usepackage{verbatim}
\usepackage{url,graphicx,tabularx,array,geometry}
\makeatletter

\def\bx{\mathbf{x}}
\def\by{\mathbf{y}}
\def\bo{\mathbf{o}}

\newcommand{\Rmnum}[1]{\expandafter\@slowromancap\romannumeral #1@}
\makeatother

\begin{document}

\title{Estimating a graphical intra-class correlation coefficient (GICC) using multivariate probit-linear mixed models}
\author{Chen YUE$^{a}$, Shaojie CHEN$^{a}$, Haris I. SAIR$^{b}$, Raag AIRAN$^{b}$, \\ Brian S. CAFFO$^{a}$\\
\footnotesize $^{a}$ Department of Biostatistics, Johns Hopkins University, Baltimore, MD, 21205\\
\footnotesize $^{b}$ Department of Radiology and Radiological Science, Johns Hopkins University, Baltimore, MD 21287\\
} 
\date{\today}
\maketitle
\begin{abstract}
Data reproducibility is a critical issue in all scientific experiments. In this manuscript, we consider the problem of quantifying the reproducibility of graphical measurements. We generalize the concept of image intra-class correlation coefficient (I2C2) and propose the concept of the graphical intra-class correlation coefficient (GICC) for such purpose. The concept of GICC is based on multivariate probit-linear mixed effect models. We will present a Markov Chain EM (MCEM) algorithm for estimating the GICC. Simulations results with varied settings are demonstrated and our method is applied to the KIRBY21 test-retest dataset.\

\


\end{abstract}

\newpage
\section{Introduction}
A crucial question in any statistical analysis is: how reliable is the data? Experimental replication for the purpose of measuring the reliability of the measurement is the most common method for establishing reproducibility. In this paper, we consider repeated measurement of graphs and propose the concept of the graphical intra-class correlation coefficient for measuring the reliability of graphs.\

The Intra-class correlation coefficient (ICC) has been proposed \cite{fisher1970statistical} and used to evaluate the reliability of measurements in a variety applications \cite{bartko1966intraclass}, \cite{shrout1979intraclass}. One way ANOVA mixed-effect models have been proposed \cite{stanish1983estimation} to estimate the ICC. Suppose $y_{ij}$ denotes the $j^{th}$ measurement of subject $i$, $x_i$ denotes the subject specific random effect and $u_{ij}$ indicates the measurement error. The one-way ANOVA model is:
\begin{equation}
\label{eq:sviccm}
\begin{split}
y_{ij} &=\mu + x_i + u_{ij}\\
x_i&\sim N(0, \sigma_x^2),\ u_{ij}\sim N(0, \sigma_u^2),\ i.i.d.
\end{split}
\end{equation}
The ICC is then defined as:
\begin{equation}
\label{eq:svicc}
ICC=\frac{\sigma_x^2}{\sigma_x^2+\sigma_u^2}.
\end{equation}
In (\ref{eq:sviccm}) and (\ref{eq:svicc}), the total variability of the data are decomposed into subject-specific variability and measurement error; ICC represents the proportion of variability that is due to heterogeneity in subjects. In recent research, the ICC has been generalized to multivariate cases. \citeauthor{di2009multilevel} \cite{di2009multilevel} proposed a model analogous to (1) in functional data using multilevel functional principal component analysis (MFPCA) and an image intra-class correlation coefficient (I2C2) was subsequently proposed in \cite{shou2012image} to calculate ICC for image data.\

Graphical data are becoming increasingly popular in scientific research. For example, graphs are used in describing brain networks in neuroimaging. In such research, binary graphs are often obtained from functional magnetic resonance image (fMRI) \cite{di2008functional}, \cite{guye2010graph}, \cite{huang2010learning}, \cite{salvador2005neurophysiological}, \cite{van2010exploring}. The increasing number of graphical datasets motivated us to evaluate the reliability of binary graphs. \

\begin{figure}[H]
\label{fig:gicc.demo}
\centering
\caption[GICC illustration]{\footnotesize{Left panels shows a high GICC case, where graphical measurements are similar within subject, right panel illustrates a low GICC case, where graphical measurements are less consistent within subject.}}
\scalebox{0.45}{\includegraphics{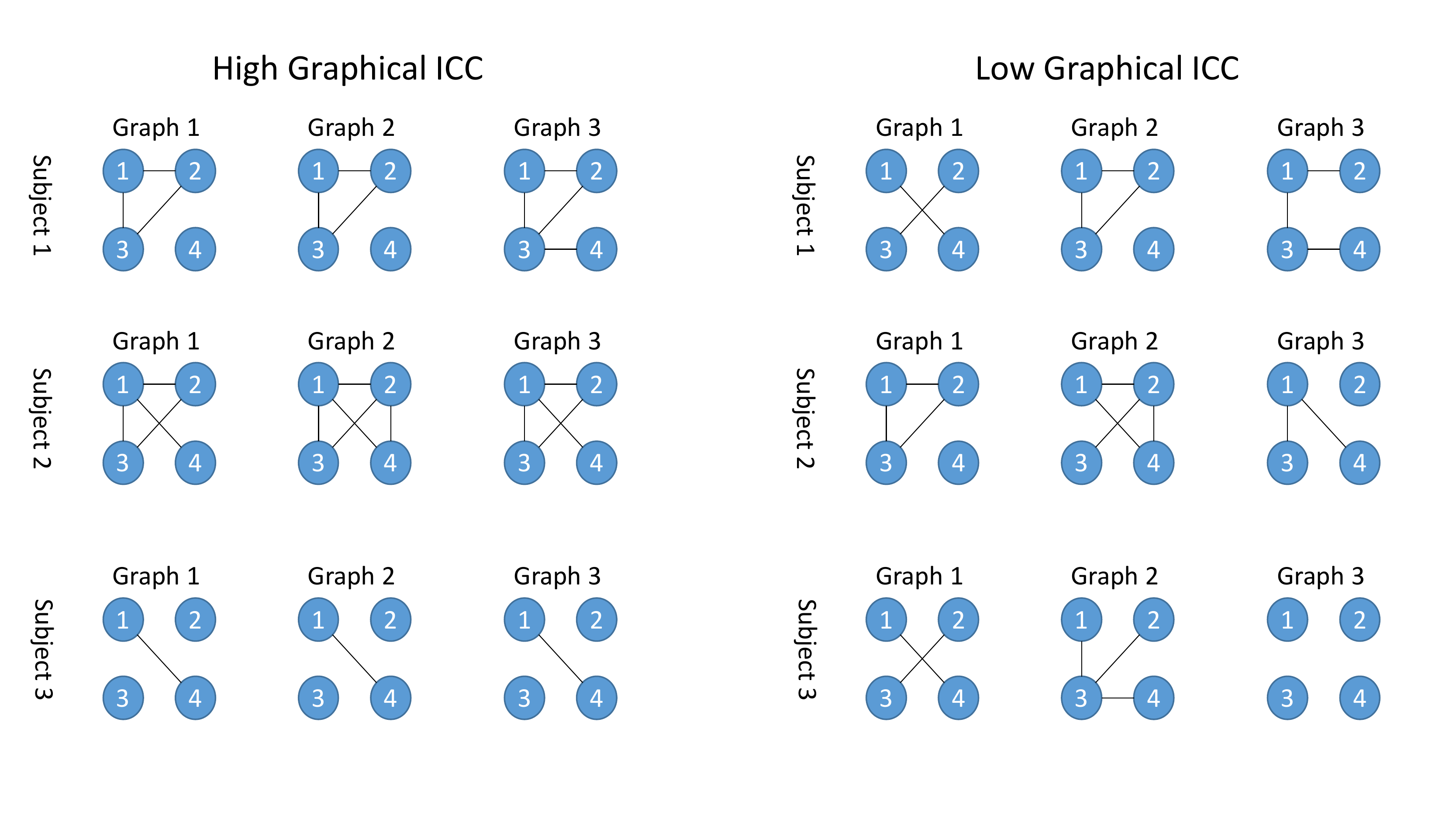}}
\end{figure}

Figure.\ref{fig:gicc.demo} illustrates idealized graphical measurements for three different subjects. Here each subject is measured three times. The left panel shows a case where graphical measurements resemble each other within one subject. The ICC, consequentially, should be higher. The right panel, on the other hand, demonstrates the opposite situation, where the repeated measurements within one subject show poor consistency. In such case, the ICC should be relatively lower. In this paper, we propose the concept of graphical ICC (GICC) to quantify the similarity between repeated measurements of binary graphs. In Figure.\ref{fig:gicc.demo}, each binary graph is represented by a $0-1$ vector. For example, the first graph of subject 1 is represented by $(1,1,0,1,0,0)^T$\footnote{Each element of the vector is an indicator of the existence of an edge, the order of the six elements is $\textcircled{1}-\textcircled{2},\ \textcircled{1}-\textcircled{3},\ \textcircled{1}-\textcircled{4},\ \textcircled{2}-\textcircled{3},\ \textcircled{2}-\textcircled{4},\ \textcircled{3}-\textcircled{4}$.}. Thus our goal is to define an ICC for multivariate binary data.\

Many authors have discussed the ICC for single variate binary data. \citeauthor{ridout1999estimating} \cite{ridout1999estimating} proposed a moment based estimator. Probit linear mixed-effect models were used by \cite{rodriguez2003intra} and \citeauthor{zou2004confidence} \cite{zou2004confidence} estimates a confidence interval for binary data ICC.\

Our objective is to estimate the ICC to evaluate the reliability of replicated measurement of binary graphs. In section \ref{sec:model}, a multivariate probit linear mixed model is proposed. Monte Carlo expectation maximization (MCEM) algorithm will be discussed in section \ref{sec:alg}. Simulation results with various settings will be shown in section \ref{sec:simul} and the results of our method being implemented on binary brain connectivity maps are in section \ref{sec:app}. We will summarize the paper in section \ref{sec:con}.

\section{Model}
\label{sec:model}
Suppose $\big\{o_{ij}(d):\ i=1,\dots,I;\ j=1\dots,J_i;\ d=1,\dots,D,\big\}$ are binary observations representing repeated graph measurements for multiple subjects. Here $I$ is the total number of subjects, $J_i$ is the number of visits for the $i^{th}$ subject and $D$ is the number of possible edges for all graphs. Usually, we have $D=\frac{N(N-1)}{2}$ where $N$ is the number of nodes. In Figure.\ref{fig:gicc.demo}, for example, we have $I=3,\ J_i=3,\ N=4,\ D=6$. The multivariate probit-linear mixed model is as follows:
\begin{equation}
\label{eq:model1}
\begin{split}
\Phi^{-1}\big(P(o_{ij}(d)|x_{i}(d))\big)&=\mu(d)+x_i(d),\\
\mathbf{x}_i&\sim\ \mathbf{N}(\mathbf{0},\mathbf{\Sigma}_x),
\end{split}
\end{equation}
where $\mathbf{x}_i=(x_i(1),\dots,x_i(D))^T$. The GICC, is then defined as:
\begin{equation}
\label{eq:GICC}
GICC=\frac{tr(\mathbf{\Sigma}_x)}{tr(\mathbf{\Sigma}_x)+D}.
\end{equation}
For the purpose of estimation, the model can also be viewed as a threshold model that dichotomizing the observations from a latent Gaussian distribution. In other words:
\begin{equation}
\begin{split}
\label{eq:model2}
o_{ij}(d)&=\mathbf{I}_{(y_{ij}(d)>0)},\\
y_{ij}(d)&=\mu(d)+x_i(d)+u_{ij}(d),\\
\mathbf{x}_i&\sim\ \mathbf{N}(\mathbf{0}, \mathbf{\Sigma}_x),\ i.i.d.,\\
\mathbf{u}_{ij}&\sim\ \mathbf{N}(\mathbf{0}, \mathbf{I}),\ i.i.d.,\\
\end{split}
\end{equation}
where $\mathbf{x}_i=(x_i(1),\dots,x_i(d))^T$ and $\mathbf{u}_{ij}=(u_{ij}(1),\dots,u_{ij}(d))^T$. The equivalency of these two models can be easily shown by the following calculation:
\begin{equation*}
\begin{split}
P(o_{ij}(d)=1|x_i(d))&=P(y_{ij}(d)>0|x_i(d))\\
&=P(u_{ij}(d)> -(\mu(d)+x_i(d))|x_i(d))\\
&=1-\Phi(-(\mu(d)+x_i(d)))\\
&=\Phi(\mu(d)+x_i(d)).\\
\end{split}
\end{equation*}

\section{The Monte Carlo EM Algorithm}
\label{sec:alg}
MCEM algorithms have been used in probit-linear mixed models with single variate outcomes \cite{chan1997maximum}. Here MCEM is generalized to the multivariate case. In model (\ref{eq:model2}), the parameters of interest are $\mathbf{\mu}$ and $\mathbf{\Sigma}_x$. In the procedure of estimation, we treat $\bo$ as observed data and $[\by,\ \bx]$ as the full data.
\subsection{M-step}
Given the full data $\by$ and $\bx$, the MLE for both parameters yields an explicit form:
\begin{equation}
\label{eq:m-step}
\begin{split}
\hat{\mathbf{\mu}}&=\frac{1}{\sum_i{J_i}}\sum_{i}\sum_{j}(\mathbf{y}_{ij}-\mathbf{x}_i),\\
\hat{\mathbf{\Sigma}_x}&=\frac{1}{I}\sum_i{\mathbf{x}_i\mathbf{x}_i^T}.\\
\end{split}
\end{equation}\

Unlike \cite{mcculloch1994maximum}, the estimate of $\mathbf{\mu}$ does not involve $\mathbf{\Sigma}_x$ since $\bx$ is also treated as part of complete data. So $\hat{\mathbf{\mu}}$ is obtained based on both $\mathbf{x}$ and $\mathbf{y}$, rather than only on $\mathbf{y}$.\

Substitute $\by$, $\bx$ and $\bx\bx^T$ with $E[\by|\bo]$, $E[\bx|\bo]$ and $E[\bx\bx^T|\bo]$ respectively on the right side of (\ref{eq:m-step}), we obtain the M-step.

\subsection{E-step}
Based on (\ref{eq:m-step}), it is necessary need to calculate $E(\by_{ij}|\bo)$, $E(\bx_i|\bo)$ and $E(\bx_i\bx_i^T|\bo)$. Since
\begin{equation}
\begin{split}
E[\mathbf{x}_i|\mathbf{o}]&=E[E[\mathbf{x}_i|\mathbf{y}]|\mathbf{o}],\\
E[\mathbf{x}_i\mathbf{x}_i^T|\mathbf{o}]&=E[E[\mathbf{x}_i\mathbf{x}_i^T|\mathbf{y}]|\mathbf{o}].
\end{split}
\end{equation}
The inner expectation can be obtained by using the joint distribution of $\big\{\mathbf{x}_i,\ \mathbf{y}_{i1},\dots,\mathbf{y}_{iJ_i}\big\}$. Noticing the following fact:
\begin{equation}
\begin{split}
\label{eq:innerexpect}
[\mathbf{x}_{i},\mathbf{y}_{i1},\dots,\mathbf{y}_{iJ_i}]=&\prod_{j}[\mathbf{y}_{ij}|\mathbf{x}_i]\times[\mathbf{x}_i]\\
\propto& \exp\Bigg\{-\frac{1}{2}\Big[\sum_j\big\{(\mathbf{y}_{ij}-\mathbf{\mu}-\mathbf{x}_i\big)^T(\mathbf{y}_{ij}-\mathbf{\mu}-\mathbf{x}_i\big)\}\\ &+\mathbf{x}_i^T\mathbf{\Sigma}_x^{-1}\mathbf{x}_i\Big]\Bigg\}\\
\propto&\exp\Bigg\{-\frac{1}{2}\Big[\mathbf{x}_i^T(J_i\mathbf{I}+\mathbf{\Sigma}_x^{-1})\mathbf{x}_i-2\big[\sum_j(\mathbf{y}_{ij}-\mathbf{\mu})\big]^T\mathbf{x}_i\Big]\Bigg\}.
\end{split}
\end{equation}
Using (\ref{eq:innerexpect}), it can be derived that:
\begin{equation}
\begin{split}
\label{eq:postdist}
\mathbf{x}_i|\mathbf{y}_{i1},\dots\mathbf{y}_{iJ_i}\sim \mathbf{N}\Big((J_i\mathbf{I}+\mathbf{\Sigma}_x^{-1})^{-1}(\mathbf{y}_{i.}-J_i\mathbf{\mu}),\ \ (J_i\mathbf{I}+\mathbf{\Sigma}_x^{-1})^{-1}\Big),
\end{split}
\end{equation}
where $\mathbf{y}_{i.}=\sum_j\mathbf{y}_{ij}$. Thus we have
\begin{equation}
\label{eq:estep}
\begin{split}
E(\mathbf{x}_i|\mathbf{O})=&E[E[\mathbf{x}_i|\mathbf{y}]|\mathbf{o}]\\
=&E\Bigg[(J_i\mathbf{I}+\mathbf{\Sigma}_x^{-1})^{-1}(\mathbf{y}_{i.}-J_i\mathbf{\mu})|\mathbf{o}\Bigg]\\
=&(J_i\mathbf{I}+\mathbf{\Sigma}_x^{-1})^{-1}\big(E[\mathbf{y}_{i.}|O]-J_i\mathbf{\mu}\big),\\
E[\mathbf{x}_i\mathbf{x}_i^T|\mathbf{o}]&=E[E[\mathbf{x}_i\mathbf{x}_i^T|\mathbf{y}]|\mathbf{o}]\\
=&E\Bigg[(J_i\mathbf{I}+\mathbf{\Sigma}_x^{-1})^{-1}(\mathbf{y}_{i.}-J_i\mu) (\mathbf{y}_{i.}-J_i\mu)^T(J_i\mathbf{I}+\mathbf{\Sigma}_x^{-1})^{-1}+(J_i\mathbf{I}+\mathbf{\Sigma}_x^{-1})^{-1}|\mathbf{o}\Bigg]\\
=&(J_i\mathbf{I}+\mathbf{\Sigma}_x^{-1})^{-1}E\big[(\mathbf{y}_{i.}-J_i\mu) (\mathbf{y}_{i.}-J_i\mu)^T|\mathbf{o}\big](J_i\mathbf{I}+\mathbf{\Sigma}_x^{-1})^{-1}+(J_i\mathbf{I}+\mathbf{\Sigma}_x^{-1})^{-1}.\\
\end{split}
\end{equation}

However, the term $E[\mathbf{y}_{i.}|\mathbf{o}]$ and $E[\mathbf{y}_{i.}^T\mathbf{y}_{i.}|\mathbf{o}]$ does not have explicit form. Here we use Gibbs sampler to approximate the conditional expectation. Notice that given $\bo$, the distribution of $\by$ is multivariate truncated normal. The Gibbs sampler for such a distribution has been discussed in \cite{horrace2005some}, \cite{kotecha1999gibbs}, \cite{tmvtnorm}. In the Gibbs sampling cycles, we choose the burn period to be the first $T=200$ and treat the following $B=500$ elements as stationary realizations coming from the conditional distribution of $\by|\bo$. Then an empirical conditional expectation could be calculated as follows:

\begin{equation}
\label{eq:MCMC}
\begin{split}
\hat{E}[\mathbf{y}_{ij}|\mathbf{o}]&=\frac{1}{B}\sum_{b=T+1}^{T+B}\mathbf{y}_{ij}^{(b)},\\
\hat{E}[\mathbf{y}_{i.}\mathbf{y}_{i.}^T|\mathbf{o}]&=\frac{1}{B}\sum_{b=T+1}^{T+B}\mathbf{y}_{i.}^{(b)}\mathbf{y}_{i.}^{(b)T}.\\
\end{split}
\end{equation}

\subsection{Observed information matrix for $\mu$}
\label{sub:obs}
Though we are not specifically interested in estimating $\mu$ for the graphical ICC, the estimate of $\mu$ with its standard error remains potential interests, especially for modeling multivariate binary data using probit-linear mixed model. \citeauthor{louis1982finding} \cite{louis1982finding} expressed the observed information matrix in EM algorithm using the first and second derivative of the full likelihood.\

Assume the observed log-likelihood is $l_o(\mathbf{\bo},\theta)$ where $\theta=(\mu, \Sigma_x)$ and the full log-likelihood is $l_{\bx,\by}(\bx,\by,\theta)$, follow \cite{louis1982finding}, we have:
\begin{equation}
\begin{split}
\label{eq:oim}
I_{\bo}(\theta)=&E_{\theta}\left[-\frac{\partial^2  l_{\bx,\by}(\bx,\by,\theta)}{\partial\mu\partial\mu^T}\Bigg|\bo\right]-E_{\theta}\left[\left(\frac{\partial l_{\bx,\by}(\bx,\by,\theta)}{\partial \mu}\right)\ \left(\frac{\partial l_{\bx,\by}(\bx,\by,\theta)}{\partial \mu}\right)^T\Bigg|\bo\right]\\
&+\left(\frac{\partial l_o(\bo,\theta)}{\partial\mu}\right)\ \left(\frac{\partial l_{\bo}(\bo,\theta)}{\partial\mu}\right)^T.
\end{split}
\end{equation}
Let $I_{\bo}=I_{\bo}(\hat{\theta})$, where $\hat{\theta}$ is the maximum likelihood estimator. Then we have:
\begin{equation}
\label{eq:oim2}
I_{\bo}(\hat{\theta})=E_{\hat{\theta}}\left[-\frac{\partial^2  l_{\bx,\by}(\bx,\by,\theta)}{\partial\mu\partial\mu^T}\Bigg|\bo\right]-E_{\hat{\theta}}\left[\left(\frac{\partial l_{\bx,\by}(\bx,\by,\theta)}{\partial \mu}\right)\ \left(\frac{\partial l_{\bx,\by}(\bx,\by,\theta)}{\partial \mu}\right)^T\Bigg|\bo\right].
\end{equation}
Following the same path in the E-step, we can use Gibbs sampler with its empirical averages to approximate the conditional expectation.

\section{Simulation}
\label{sec:simul}
We set number of subject at $I=100,\ 200$ and each subject receives $J=2,\ 4$ repeated measurements. The number of nodes is set to be $N=5$ so that the number of possible edges is $D=10$. The true $\mu$ is set to be 0.5 for all elements and $$\Sigma_x[i,j]=r\rho^{|i-j|},\ \text{where}\ \rho=0.8.$$ The underlying true graphical ICC using definition (\ref{eq:GICC}) is controlled by $r$. We did $r=2,\ 4$ in each settings so that the corresponding ICC's are $\frac{rD}{rD+D}=2/3,\ 4/5$ respectively. A total of 500 simulations were run in each simulation group.\

In Table. \ref{tab:simulation}, the average estimated GICC for $r=2$ groups are 0.702, 0.672, 0.683 for $I_{100}J_{2}$, $I_{100}J_{4}$ and $I_{200}J_{4}$ group respectively, comparing to an underlying truth $2/3\approx 0.667$. As number of individuals increases, or as number of repeated measurements increases, both bias and standard deviation of the estimated GICC reduces. When $r=4$, the average estimated graphical ICC are 0.817, 0.800 and 0.806 respectively. The MLE of GICC in each case has a positive bias, which is reduced as either $I$ or $J_i$ increases.

\section{Application}
\label{sec:app}

Resting-state fMRI scans consisted of a test-retest dataset previously acquired at the FM Kirby Research Center at the Kennedy Krieger Institute, Johns Hopkins University \cite{landman2011multi}. Twenty one healthy volunteers with no history of neurological disease each underwent two separate resting state fMRI sessions on the same scanner. A 3T MR scanner was used (Achieva, Philips Healthcare, Best, The Netherlands) utilizing a body coil with a 2D echoplanar (EPI) sequence and eight channel phased array SENSitivity Encoding (SENSE; factor of 2) with the following parameters: TR 2s, 3mm x 3mm in plane resolution, slice gap 1mm, for total imaging time of 7 minutes and 14 seconds. One subject was excluded due to technical issues at acquisition.

ICA (Independent Component Analysis) was performed using MEDOLIC (Multivariate Exploratory Linear Optimized Decomposition into Independent Components) version 3.10 in FSL (FMRIB Software Library, FMRIB, Oxford, UK). Preprocessing included removal of low-frequency drift with a highpass filter cutoff of 250s, realignment of the fMRI time series using MCFLIRT, slice timing correction, brain extraction using BET, and spatial smoothing with FWHM of 6mm. Images were registered to MNI standard space with resampling resolution of 2mm. ICA was performed using multi-session temporal concatenation with automatic dimensionality estimation and time-course variance normalization implemented in MELODIC. 43 components were identified by MELODIC.

Relevant ICA components corresponding to known large scale brain networks were identified by a board certified neuroradiologist with experience in resting state fMRI. Seven total components were selected (default mode network, dorsal attention network, motor network, visual network, salience network, and two lateralized executive control networks), and the 7 by 7 correlation matrix was calculated (see raw data in Figure 2). Different thresholds were used to dichotomize the raw graphs into binary ones, where the thresholds were chosen from 0.1 to 0.8 using grid 0.01 (see Figure 2). The GICC algorithm was then implemented on these binary graphs.

The GICC was then calculated for each threshold (see Figure 3). The GICC remains above 0.6 when the threshold is between 0.1 and 0.6. For threshold outside of this band the GICC decreases dramatically. When threshold is around 0.8, GICC fluctuates more significantly and the value eventually drops to 0.1. Thus the GICC shows high reproducibility of the raw data if a reasonable threshold is employed (from 0.1 to 0.6). When the threshold is too high, only few raw values will be dichotomize to 1 such that poor reproducibility is obtained. For practical subsequent application, one could use the value that maximizes the GICC in this data set, which, in this case, is 0.35 (see Figure 3).

%

\section{Conclusion}
\label{sec:con}

In this paper, we propose the concept of graphical intra-class correlation coefficient using multivariate probit mixed-linear model. The GICC is defined as $\frac{tr(\Sigma_x)}{tr(\Sigma_x)+D}$. We use Monte Carlo EM algorithm to obtain the MLE of $\Sigma_x$ while Gibbs sampler was used in the E-step. We show the results of GICC in varied simulation settings and in the KIRBY21 test-retest datasets. \

While providing GICC, the estimation procedure can also be generalized to multivariate probit mixed-linear model with fixed and random covariates components, which is:
\begin{equation*}
\begin{split}
\label{eq:model2}
o_{ij}(d)&=\mathbf{I}_{(y_{ij}(d)>0)}\\
y_{ij}(d)&=\sum_{p=1}^P\mu_{ip}(d)\beta_p(d)+\sum_{r=1}^{R}x_{ir}(d)\eta_{ir}(d)+u_{ij}(d)\\
\mathbf{\eta}_{ir}&\sim\ \mathbf{N}(\mathbf{0}, \mathbf{\Sigma}_{r}),\ i.i.d.\\
\mathbf{u}_{ij}&\sim\ \mathbf{N}(\mathbf{0}, \mathbf{I}),\ i.i.d..\\
\end{split}
\end{equation*}
In the EM algorithm, $\mathbf{\eta}$ and $\by$ can be treated as full data and the procedure in Section \ref{sec:alg} follows. In section \ref{sub:obs}, we also calculate the observed information matrix for the fixed effects which can provide confidence intervals for $\mathbf{\beta}$'s. Moreover, the procedure can also be used multivariate generalized mixed-linear models, such as multivariate poisson or logistic regression.\

Currently, our method works for small graphs. As the number of nodes in a graph increase, the number of parameters of interests grows quartically ($D\sim O(N^2)$ and $\#\{\sigma_{ij}\}\sim O(D^2)$). Thus a graph with 100 nodes will have approximately $1/4\times10^8$ parameters to be estimate. The Gibbs sampler could not be implemented effectively in such cases. Therefore, the algorithm currently requires a relative small number of nodes for each graph (typically less than 10). In order to achieve faster convergence rate as well as control the Monte Carlo error induced by Gibbs sampler, ascent based MCEM \cite{caffo2005ascent} and accelerating EM algorithm \cite{varadhan2008simple} could be implemented in the MCEM.\

Notice that from the application, GICC could also be used for choosing threshold for dichotomizing raw graphs. The value that maximizes the GICC is a reasonable threshold, since it yields the best reproducibility of a well known benchmark data set.\

In summary, GICC provides us a way to measure the reproducibility of repeated graphical measurements. The current algorithm gives us the estimates of GICC for relatively small graphs. To our knowledge, GICC for large graphs has not been addressed before and therefore deserves further investigation.\

\newpage
\begin{appendix}
\listoffigures
\listoftables

\newpage
\section{Figures}
\

\newpage
\begin{figure}[H]
\label{fig:app}
\centering
\caption[Data illustration]{\footnotesize{The figure illustrates two repeated measurements for one subject. On the left, raw correlations between seven nodes are illustrated. Then the raw correlations are dichotomizing using different thresholds (0.2, 0.35, 0.6 are listed here). Our algorithm is then implemented on binary graphes using each threshold. Red suggests lower value and white (yellow) suggests higher value. In the binary graph on the right, red indicates 1 and yellow indicates 0.}}
\hspace{-2cm}
\scalebox{0.5}{\includegraphics{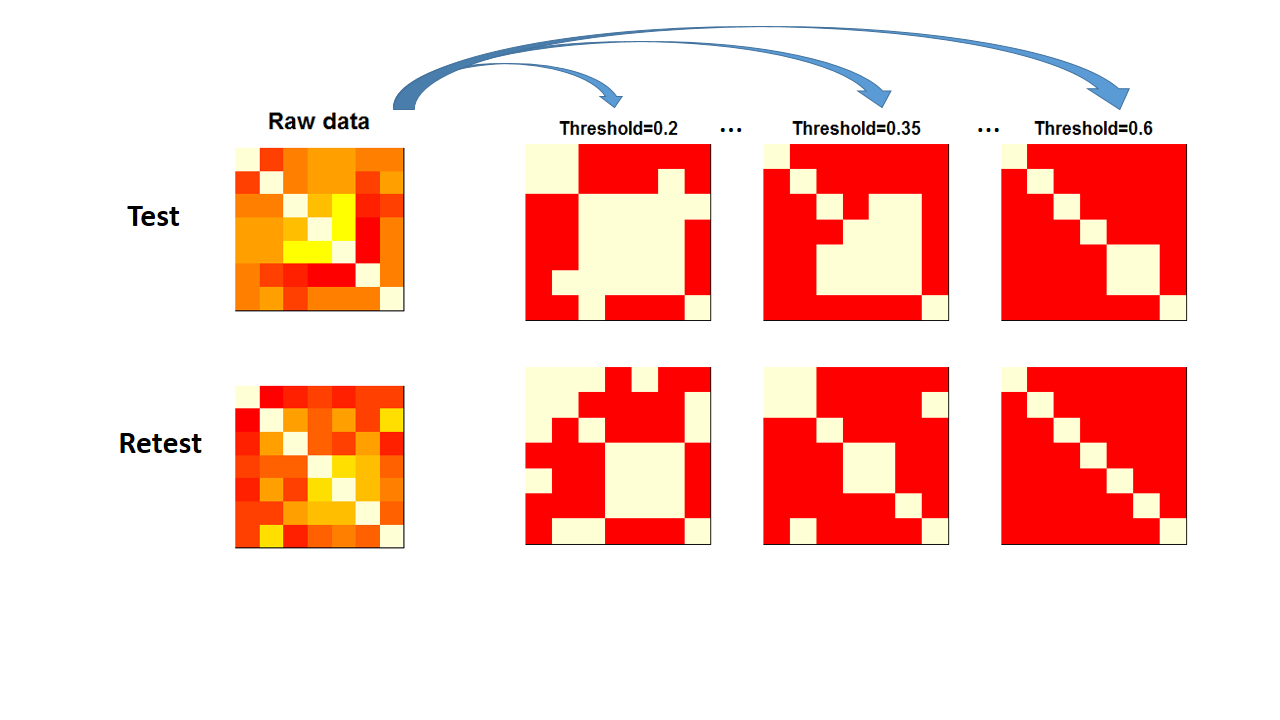}}
\end{figure}\

\newpage
\begin{figure}[H]
\label{fig:app2}
\caption[GICC curve]{\footnotesize{The calculated GICC under different thresholds. The threshold were picked equally spaced from 0.1 to 0.8 using grid 0.01. The maximized GICC is indicated in the figure, which corresponds to a 0.35 threshold.}}
\scalebox{0.7}{\includegraphics{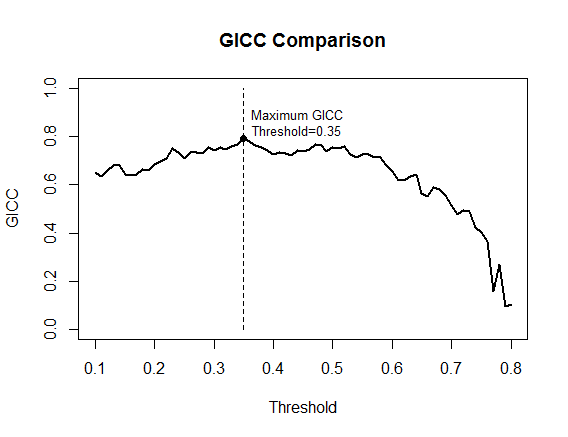}}
\end{figure}\

\newpage
\section{Tables}
\renewcommand{\arraystretch}{0.5}
\begin{table}[H]
\label{tab:simulation}
\caption[Simulation results]{Simulation results}
\centering
\begin{tabular}{lcccccc}
Setting & \multicolumn{5}{c}{Estimates for $\sigma_{ii}$} & ICC est.\\
\hline\hline
\multirow{3}{*}{$I=100,\ J=2,\ r=2$ } & $\sigma_{1,1}$ & $\sigma_{2,2}$ & $\sigma_{3,3}$ & $ \sigma_{4,4}$ & $\sigma_{5,5}$ & $ICC$\\
\cline{2-6}
& 2.37 & 2.48 & 2.38 & 2.35 & 2.34 & \multirow{2}{*}{0.702}\\
& (1.18) & (1.09) & (1.11) & (1.04) & (1.01) &\\
\cline{2-6}
\multirow{3}{*}{$ICC_{true}=2/3$ } & $\sigma_{6,6}$ & $\sigma_{7,7}$ & $\sigma_{8,8}$ & $ \sigma_{9,9}$ & $\sigma_{10,10}$ & \\
\cline{2-6}
& 2.45 & 2.36 & 2.43 & 2.43 & 2.38 &\multirow{2}{*}{(0.033)}\\
& (1.09) & (1.06) & (1.13) & (1.04) & (1.18) &\\
\hline\hline
\multirow{3}{*}{$I=100,\ J=2,\ r=4$ } & $\sigma_{1,1}$ & $\sigma_{2,2}$ & $\sigma_{3,3}$ & $ \sigma_{4,4}$ & $\sigma_{5,5}$ & $ICC$\\
\cline{2-6}
& 4.56 &4.74 &4.62 &4.40 &4.56 & \multirow{2}{*}{0.817}\\
& (2.11) & (2.20) & (2.30) & (2.01) & (2.28) &\\
\cline{2-6}
\multirow{3}{*}{$ICC_{true}=4/5$ } & $\sigma_{6,6}$ & $\sigma_{7,7}$ & $\sigma_{8,8}$ & $ \sigma_{9,9}$ & $\sigma_{10,10}$ & \\
\cline{2-6}
& 4.65 & 4.63 & 4.52 & 4.61 & 4.51 &\multirow{2}{*}{(0.025)}\\
& (2.12) & (2.04) & (2.07) & (2.04) & (2.31) &\\
\hline\hline
\multirow{3}{*}{$I=100,\ J=4,\ r=2$ } & $\sigma_{1,1}$ & $\sigma_{2,2}$ & $\sigma_{3,3}$ & $ \sigma_{4,4}$ & $\sigma_{5,5}$ & $ICC$\\
\cline{2-6}
& 2.02 & 2.10 & 2.07 & 2.04 & 2.10 & \multirow{2}{*}{0.672}\\
& (0.60) & (0.64) & (0.61) & (0.65) & (0.65) &\\
\cline{2-6}
\multirow{3}{*}{$ICC_{true}=2/3$ } & $\sigma_{6,6}$ & $\sigma_{7,7}$ & $\sigma_{8,8}$ & $ \sigma_{9,9}$ & $\sigma_{10,10}$ & \\
\cline{2-6}
& 2.08 & 2.06 & 2.08 & 2.08 & 2.05 &\multirow{2}{*}{(0.026)}\\
& (0.59) & (0.60) & (0.63) & (0.65) & (0.61) &\\
\hline\hline
\multirow{3}{*}{$I=100,\ J=4,\ r=4$ } & $\sigma_{1,1}$ & $\sigma_{2,2}$ & $\sigma_{3,3}$ & $ \sigma_{4,4}$ & $\sigma_{5,5}$ & $ICC$\\
\cline{2-6}
& 4.00 & 4.08 & 4.04 & 3.96 & 4.17 & \multirow{2}{*}{0.800}\\
& (1.29) & (1.36) & (1.28) & (1.29) & (1.36) &\\
\cline{2-6}
\multirow{3}{*}{$ICC_{true}=4/5$ } & $\sigma_{6,6}$ & $\sigma_{7,7}$ & $\sigma_{8,8}$ & $ \sigma_{9,9}$ & $\sigma_{10,10}$ & \\
\cline{2-6}
& 4.04 & 4.04 & 4.10 & 4.09 & 4.05 &\multirow{2}{*}{(0.020)}\\
& (1.24) & (1.20) & (1.34) & (1.32) & (1.24) &\\
\hline\hline
\multirow{3}{*}{$I=200,\ J=2,\ r=2$ } & $\sigma_{1,1}$ & $\sigma_{2,2}$ & $\sigma_{3,3}$ & $ \sigma_{4,4}$ & $\sigma_{5,5}$ & $ICC$\\
\cline{2-6}
& 2.08 & 2.22 & 2.23 & 2.17 & 2.19 & \multirow{2}{*}{0.683}\\
& (0.68) & (0.77) & (0.74) & (0.73) & (0.70) &\\
\cline{2-6}
\multirow{3}{*}{$ICC_{true}=2/3$ } & $\sigma_{6,6}$ & $\sigma_{7,7}$ & $\sigma_{8,8}$ & $ \sigma_{9,9}$ & $\sigma_{10,10}$ & \\
\cline{2-6}
& 2.17 & 2.17 & 2.16 & 2.20 & 2.13 &\multirow{2}{*}{(0.026)}\\
& (0.76) & (0.68) & (0.71) & (0.77) & (0.74) &\\
\hline\hline
\multirow{3}{*}{$I=200,\ J=2,\ r=4$ } & $\sigma_{1,1}$ & $\sigma_{2,2}$ & $\sigma_{3,3}$ & $ \sigma_{4,4}$ & $\sigma_{5,5}$ & $ICC$\\
\cline{2-6}
& 4.07 & 4.33 & 4.25 & 4.20 & 4.18 & \multirow{2}{*}{0.806}\\
& (1.38) & (1.41) & (1.57) & (1.49) & (1.42) &\\
\cline{2-6}
\multirow{3}{*}{$ICC_{true}=4/5$ } & $\sigma_{6,6}$ & $\sigma_{7,7}$ & $\sigma_{8,8}$ & $ \sigma_{9,9}$ & $\sigma_{10,10}$ & \\
\cline{2-6}
& 4.20 & 4.20 & 4.25 & 4.30 & 4.15 &\multirow{2}{*}{(0.020)}\\
& (1.52) & (1.36) & (1.43) & (1.47) & (1.47) &\\
\hline\hline
\end{tabular}
\end{table}\

\bibliographystyle{plainnat}
\nocite{*}
\bibliography{gicc_cite}
\end{appendix}
\end{document}